\documentclass[lettersize,journal]{IEEEtran}
\usepackage{amsmath,amsfonts}
\usepackage{algorithmic}
\usepackage{algorithm}
\usepackage{array}
\usepackage[caption=false,font=normalsize,labelfont=sf,textfont=sf]{subfig}
\usepackage{textcomp}
\usepackage{stfloats}
\usepackage{url}
\usepackage{verbatim}
\usepackage{graphicx}
\usepackage{cite}
\usepackage{subfloat}
\usepackage{makecell}
\usepackage{setspace}
\usepackage{lineno}

\begin{document}

\title{New  Results for the Pointing Errors Model in Two Asymptotic Cases}

\author{Maoke~Miao,
        Xiao-yu~Chen,
        Rui Yin,
        and Jiantao Yuan
\thanks{This work is supported by the National Natural Science Foundation of China (Grant No.61871347) and  Scientific Research Foundation of Zhejiang University City College (No. J-202321). \\
\indent Maoke~Miao is with the School of Information Science and Electronic Engineering, Zhejiang University, and the School of Information and Electrical Engineering, Hangzhou City University, Hangzhou, China (e-mail: maokemiao@zju.edu.cn). \\
 \indent Xiao-yu~Chen, Rui Yin, and Jiantao Yuan are  with the School of Information and Electrical Engineering, Hangzhou City University, Hangzhou 310015, China (chenxiaoyu@zucc.edu.cn; yinrui@zucc.edu.cn; yuanjt@zucc.edu.cn).
}}

\markboth{IEEE PHOTONICS JOURNAL}%
{Shell \MakeLowercase{\textit{et al.}}: A Sample Article Using IEEEtran.cls for IEEE Journals}


\maketitle

\begin{abstract}
Several precise and computationally efficient results for pointing errors models in two asymptotic cases are derived in this paper. The normalized mean-squared error (NMSE) performance metric  is employed to quantify the accuracy of different models. For the case that the beam width is relatively larger than the detection aperture, we propose the three kinds of models that have the form of $c_1 \exp\left(-c_2r^2\right)$. It is shown that the modified intensity uniform model not only achieves a  comparable accuracy with the best linearized model, but also is expressed in an elegant mathematical way when compared to the traditional Farid model. This indicates that the modified intensity uniform model is preferable in the performance analysis of free space optical (FSO) systems  considering the effects of the pointing errors. By analogizing the beam spot with a point in the
case that beam width is smaller than the detection aperture, the solution of the pointing errors model  is transformed to
a smooth function approximation problem, and we find that a more accurate  approximation can be achieved by the proposed point approximation model when compared to the  model that is induced from the Vasylyev model in some scenarios.
\end{abstract}


\section{Introduction}
\label{sec:introduction}
\indent \IEEEPARstart{F}{ree}-space optical (FSO) is a wireless optical communication technology, which has attracted considerable attention in both academic and industry  due to its great potential: larger bandwidth and
high data rate, unregulated spectrum, low mass and less power requirements, rapid and easy deployment \cite{Shao22}. Also, FSO technology can be deployed together in the so-called hybrid radio frequency (RF)/FSO systems that are considered to be a promising solution for reliable wireless backhaul connectivity to enable long-range communications in future  6G  networks \cite{Alsabah21}. Furthermore, the advanced reconfigurable intelligent surfaces (RIS) technology that was proposed recently, can be used to enhance the performance of FSO systems, and thus broadening the range of FSO communication \cite{Najafi21}. \\
\indent  Despite these benefits of FSO technology,  the performance of FSO communication systems can be deteriorated by adverse effects, such as beam wandering and spreading, and scatting  when the optical carrier propagates through  atmospheric turbulence. It has been shown in \cite{Kaushal17}  that turbulence-induced irradiance  scintillation and pointing errors are the two major performance-limiting factors for FSO links with ranges longer than one kilometer.  Note that the beam wander and mechanical vibration result in the pointing errors, and they have the same mathematical model with only differing in physical meaning of parameters \cite{Trichili20,KAUSHAL2017106}. The research on the irradiance scintillation models have been studied extensively, and plenty of precise or mathematically tractable  models have been proposed so far, such as Gamma-Gamma distribution \cite{Andrews05}, Fischer-Snedecor $\mathcal{F}$ distribution \cite{Peppas20},  lognormal-Rician distribution \cite{Miao20}, and  {Málaga distribution} \cite{Ansari16}. Unfortunately, the results  for the pointing errors models are greatly limited. To the best of the author's knowledge, the  pointing errors model for a  Gaussian beam was firstly developed by R. Esposito   in \cite{Esposito67}, where it was expressed in terms of Marcum's Q function\footnote{Marcum's Q function plays an important role in the performance analysis of communication systems, which is defined as $Q(\alpha, \beta) = \int_\beta^{\infty} t \exp \left(-{(t^2+x^2)}/{2}\right) I_0(\alpha t) dt$ \cite{RM339PR}.}. Subsequently, a simple and efficient approximation of this model was present by Farid
 \cite{Farid07}, namely, Farid model, which has been widely used in FSO systems. It should be noted, however, the Farid  model has two main drawbacks: 1) one is the low approximation accuracy  when the radius of beam width is  two times less than that of the detection aperture. 2) the other is that it requires the computation of the complex error function $\text{erf}\left(\cdot\right)$. Recently, the other
pointing errors model was established by Vasylyev in the field of quantum communication  \cite{Vasylyev12}. Although it provides a good approximation regardless of the relationship between   beam width and detection aperture, its complicated mathematical form greatly hampers the analytic expression for the  system performance.  \\
\indent In this work, we present some new results on the pointing errors model.  Several computation-efficient models are proposed, and the accuracy of them is investigated in detail. For the case that the beam width  is relatively larger than the detection aperture, the normalized mean-squared error (NMSE) performance  indicates that  the proposed modified intensity uniform model not only shows a better approximation accuracy than the Farid model, but also has a simpler  expression. This is one of the key contributions of this paper. By analogizing the beam spot with a point in the case that beam width is smaller than the detection aperture, the  solution of  the
 pointing errors model  is transformed to a smooth function approximation problem. Numerical results demonstrate that the  proposed  point approximation model  provides a higher accuracy  than the model induced from the Vasylyev model in some scenarios.

\section{Pointing Errors Model}
\indent In line-of-sight FSO communication links, misalignment between transmitter and receiver results in pointing errors, which are the another performance-limiting factor besides turbulence-induced scintillation. We note that the pointing errors consist  of two parts in practical FSO systems. One is the beam wandering caused by large scale eddy and the other is due to mechanical vibration or  thermal expansion. However, the former can be dealt with in a similar methodology to the latter, as shown in \cite{Trichili20,KAUSHAL2017106}.\\
\indent After propagating a distance $z$ from the transmitter, the normalized spatial distribution of a Gaussian beam  at the receiver plane is given by
\begin{equation}\label{eqp1}
  I_{\text {beam }}(\boldsymbol{\rho} ; z)=\frac{2}{\pi w_z^2} \exp \left(-\frac{2\|\boldsymbol{\rho}\|^2}{w_z^2}\right),
\end{equation}
where $\boldsymbol{\rho}$ denotes the displacement from the beam center. According to \cite{Saleh244535}, the beam radius $w_z$  at the distance $z$ is related to the beam waist $w_0$ at $z = 0$, wavelength $\lambda$, and atmospheric coherence length $\rho_0$, which can be expressed as
\begin{equation}\label{eqp2}
  w_z \approx w_o\left[1+\varepsilon\left(\frac{\lambda z}{\pi w_o^2}\right)^2\right]^{\frac{1}{2}},
\end{equation}
where $\varepsilon=\left(1+2 w_o^2 / \rho_o^2(z)\right)$. Specifically,  $\rho_0  = \left(0.55C_n^2k^2z\right)^{-3/5}$ for the  spherical wave with $C_n^2$ and $k = 2\pi/\lambda$ denoting the  index of refraction
structure constant and wave number respectively.
\begin{figure}
  \centering
  \includegraphics[width=2.5in]{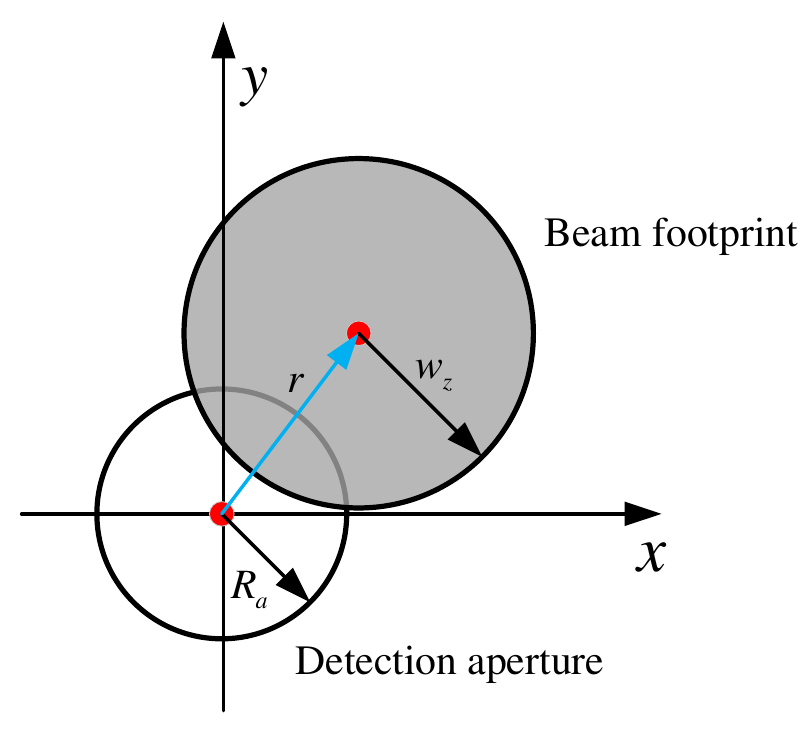}
  \caption{Schematic description of the deflection distance $r$ between the beam center and the detector center.}\label{Fig1}
\end{figure}

\indent At the receiver, the effect of pointing errors causes the deflection between the beam center and aperture center, as shown in  Fig.~\ref{Fig1}. Hence, the transmission efficiency within a circular detection aperture of radius $R_a$ reads as
\begin{equation}\label{eqp3}
  h_{{p}}(\boldsymbol{r} ; z)=\int_{\mathcal{A}} I_{\text {beam }}(\boldsymbol{\rho}-\boldsymbol{r} ; z) d \boldsymbol{\rho},
\end{equation}
where $\mathcal{A}$ is the detector area. Considering the symmetry of the beam shape and the detector area, $h_{{p}}(\boldsymbol{r} ; z)$ depends on the radial distance $r = ||\boldsymbol{r}||$, which is given by
\begin{equation}\label{eqp4}
  h_{{p}}(r ; z)=\int_{-R_a}^{R_a} \int_{-\zeta}^\zeta \frac{2}{\pi w_z^2} \exp \left(-2 \frac{\left(x^{}-r\right)^2+y^{ 2}}{w_z^2}\right) d y d x,
\end{equation}
where $\zeta=\sqrt{R_a^2-x^{ 2}}$. Equivalently, (\ref{eqp4}) can be expressed in terms of the incomplete Weber Integral, which is found to be \cite{Vasylyev12,alma71}
\begin{equation}\label{eqp5}
\begin{split}
  h_{{p}}(r ; z)&=\frac{4}{ w_z^2} \exp\left({-2 \frac{r^2}{w_z^2}}\right) \int_0^{R_a}  \varrho \exp\left({-2 \frac{\rho^2}{w_z^2}} \right) \\
  & \times {I}_0\left(\frac{4}{w_z^2} r \varrho\right)\mathrm{~d} \varrho,
\end{split}
\end{equation}
where $I_n\left(\cdot\right)$ is the modified Bessel function. From (\ref{eqp5}),  the pointing errors model at $r = 0$ can be easily derived as
\begin{equation}\label{eqp5Add1}
  h_p\left(r=0;z\right) = 1 - \exp\left(-2\frac{R_a^2}{w_z^2}\right).
\end{equation}
\section{New Results of Pointing Errors Model In Two Asymptotic Cases}
\indent In this section, we provide several methods to evaluate the pointing errors model in two asymptotic cases: $w_z \gg R_a$ and $R_a \gg w_z$. In most current  practical FSO systems, the divergence of emitted  laser beam is typical  of tens of $\mu$rad while the size of the receiving aperture is on the order of tens of centimeters \cite{Miao22,Singh22}. Hence, these two scenarios can occur, depending on the transmitted distance $z$.
\subsection{Models for $w_z \gg R_a$.}
\begin{figure}
  \centering
  \includegraphics[width=2.5in]{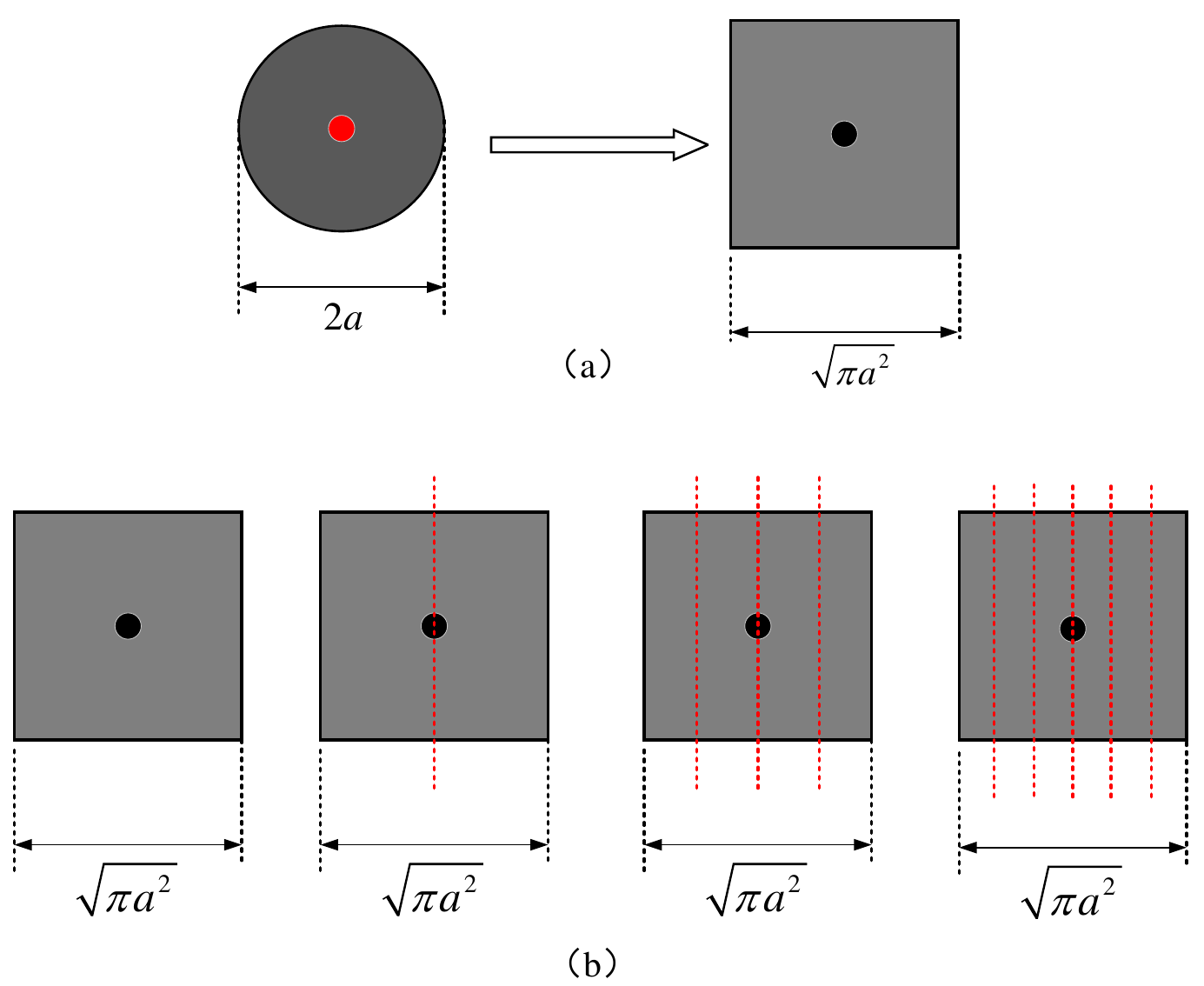}
  \caption{Two parts of Linearized Model. (a) Transformation of equal area
 between the circle and the square. (b) Partition: Examples of 0, 2, 4, 6 partitions are shown from left to right respectively. }\label{Fig2}
\end{figure}

\indent We demonstrate that the expression for pointing errors has the form $h_p\left(r;z\right) = c_1 \exp\left(-c_2 r^2\right)$ by combing the condition $w_z \gg R_a$ and (\ref{eqp1}). Note that this simple form has the benefit to facilitate the performance analysis of FSO systems provided that the radial displacement $r$ follows a Rayleigh distribution, which is given by
\begin{equation}\label{Aeq1Add1}
  f_r(r)=\frac{r}{\sigma_s^2} \exp \left(-\frac{r^2}{2 \sigma_s^2}\right), \quad r>0
\end{equation}
  where $\sigma_s^2$ is the  jitter variance at the receiver. With (\ref{Aeq1Add1}),  the unified  probability density function (PDF) of $h_p$ is obtained as
\begin{equation}\label{Aeq1Add2}
  f_{h_p}\left(h_p\right)=\frac{\gamma^2}{c_1^{\gamma^2}} h^{\gamma^2-1},
\end{equation}  
  where $\gamma^2 = 1 / \left(2\sigma_s^2 c_2\right)$. In what follows, we aim at  determining the values of coefficients $c_1$ and $c_2$ for different pointing errors models. It should be emphasized that the present results in the following are expressed in terms of elementary functions, avoiding the computation of complicated function, that is, $\text{erfc}\left(\cdot\right)$ and $I_n\left(\cdot\right)$  in Farid model \cite[eqn.~(9)]{Farid07} and Vasylyev model \cite[eqn.~(D3)]{Vasylyev12} respectively.\\
\indent  \emph{1) Intensity Uniform Model:} It can be reasonably  claimed that the intensity distribution within the area of detector aperture is approximately uniform when $w_z \gg R_a$, and we can regard the intensity of detector center as an intensity value of this area. As such, we have
\begin{equation}\label{Aeq1}
  h_p\left(r;z\right) = \pi R_a^2 I_{\text {beam }}(\boldsymbol{r} ; z) = \frac{2R_a^2}{w_z^2}\exp\left(-\frac{2r^2}{w_z^2}\right).
\end{equation}
In this case, $c_1 = {2R_a^2}/{w_z^2}$ and $c_2 = 2/w_z^2$. \\
\indent \emph{2) Modified  Intensity Uniform Model:} Although the expression for the {intensity uniform model}  is simple, it  leaves out some important details. For example, $c_1 = h_p\left(r = 0;z\right) = \eta = 1 - \exp\left(-2 R_a^2 / w_z^2\right)$ according to (\ref{eqp5}) while it is $2R_a^2 / w_z^2$ in {intensity uniform model}. Note that $2R_a^2 / w_z^2$ is a Taylor series approximation of $1 - \exp\left(-2 R_a^2 / w_z^2\right)$. Inspired by this result,  coefficient $c_2$ in (\ref{Aeq1}) may exhibit  the  same behaviour  as the coefficient $c_1$, that is, $2 / w_z^2$ is a Taylor approximation of some function. Specifically,  we have
\begin{equation}\label{Aeq2}
  h_p\left(r;z\right) = \eta \exp\left(-\eta\frac{r^2}{R_a^2}\right).
\end{equation}
when an exponential function is considered. From (\ref{Aeq2}), we have $c_1 = \eta$ and $c_2 = \eta / R_a^2$. \\
\indent \emph{3) Linearized Model:}  The coefficient $c_1$ in {linearized model} is $ 1 - \exp\left(-2 R_a^2 / w_z^2\right)$, which is the same as that in {modified  intensity uniform model}, and  the coefficient $c_2$ is determined in another way. Fig.~\ref{Fig2} depicts the process of solving the coefficient $c_2$, which consists of two steps: the circle-square transformation, and the equal space partition\footnote{The reason why we only carry out circular-square transformation and equal space segmentation are as follows: 1) (4) is difficult to solve analytically since the upper and lower limits of integration for the x- and y- axes satisfy the equation of a circle. However, the idea of ``circular-square transformation"  can be used to remove this relationship between them. 2) The ``equal space segmentation" is inspired by the specific “uniform partitions” in definite integral, which is the simplest method of  partitioning.}. \\
\begin{figure}
  \centering
  \includegraphics[width=2.5in]{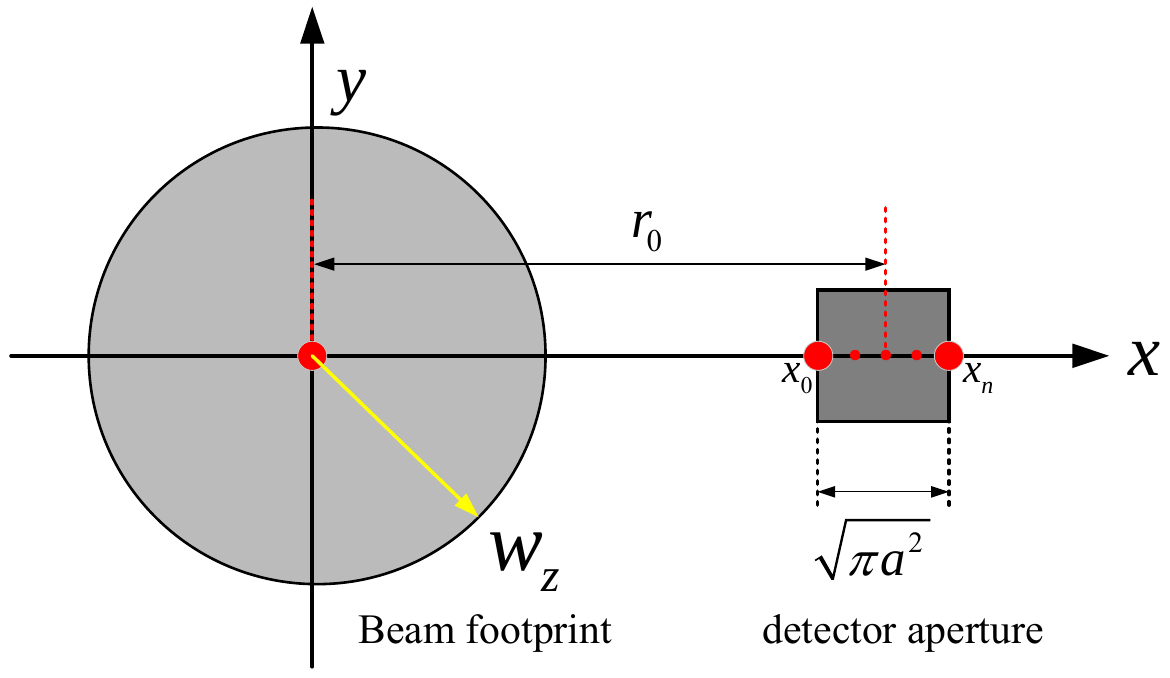}
  \caption{Equal space partition along the $x$-axis.}\label{Fig3}
\end{figure}
\indent To calculate the $c_2$ in this model, we assume that the equal space partition step only operates along one axis, and the intensity in each interval forms a linear relationship with the values from this axis while keeping the intensity  same in the other axis. Specifically, the obtained result is independent of the axis due to the symmetry, and an example of the equal space partition along the $x$-axis is shown in  Fig.~\ref{Fig3}, where   $x_n - x_0 = n \delta = \sqrt{\pi a^2}
$ with $n$ and $\delta$ denoting the number  of splits and spacing respectively. \\
\indent Hence, based on the above description, the intensity distribution satisfies the following relation
\begin{equation}\label{Aeq3}
  I_{\text {beam }}(x, y ; z) \approx I_{\text {beam }}(x ; z)=k_i x+b_i, x \in\left[x_i, x_{i+1}\right),
\end{equation}
 where $x_{i+1} = x_i + \delta$, $k_i, b_i$ are the coefficients of a linear function in the $i$-th interval, and they can be determined by  two distant points, i.e., $\left(x_i,I_{\text{beam}}\left(x_i;z\right)\right)$ and $\left(x_{i+1},I_{\text{beam}}\left(x_{i+1};z\right)\right)$. Then, the pointing errors model $h_p \left
(r;z\right)$ is approximated as
\begin{equation}\label{Aeq4}
\begin{split}
  h_p\left
(r;z\right) &\approx \sum  \limits_{i=0}^{n-1} \int_{-\sqrt{\pi R_a^2 / 4}}^{\sqrt{\pi R_a^2 / 4}} \int_{x_i}^{x_{i+1}}\left(k_i x+b_i\right) d x dy\\
& \approx \sum  \limits_{i=0}^{n-1} {\sqrt{\pi R_a^2/4}}\delta\left(k_i\delta +  2x_i\delta + 2b_i\right).
\end{split}
\end{equation}
\indent The detailed description of the procedure for the determination of $c_2$ in {linearized model} is provided  below\footnote{Note that  the radial distance $r_0$ can be optimized to improve the approximation accuracy of the linearized model. In   the numerical section, we show that the normalized optimized parameter $r_0^*/R_a$ has a quadratic relation with  $w_z/R_a$, and this indicates that the number of inputs to  this model can be reduced by one.}.

\begin{algorithm}
\caption{Algorithm for the determination of $c_2$ in \emph{Linearized Model}}
\label{alg1}
\begin{algorithmic}[1]
 \REQUIRE  Radius of detector aperture $R_a$, beam radius $w_z$, number of splits $n$,  radial distance $r_0$.
 \ENSURE  The coefficient $c_2$. \\
  \textit{Initialisation}: $l = \sqrt{\pi R_a^2}$, $Q = 0$, $c_1 = 1 - \exp\left(-2R_a^2 / w_z^2\right)$, $\delta = l/n$.
  \FOR {$i = 0$ to $n-1$}
  \STATE $k_i= \big(I_{\text {beam }}(r_0-l / 2+\delta(i+1);z)
        -I_{\text {beam }}(r_0-l / 2+\delta i;z)\big) / \delta$.
  \STATE $b_i=I_{\text {beam }}(r_0-l / 2+\delta i;z)-k_i(r_0-l / 2+\delta i)$.
  \STATE $Q = Q + {\sqrt{\pi R_a^2/4}}\delta\left(k_i\delta +  2x_i\delta + 2b_i\right)$.
  \ENDFOR
  \\
  $c_2 = -\ln\left(Q / c_1\right) / r_0^2$.
 \RETURN $c_2$
 \end{algorithmic}
 \end{algorithm}
\indent \emph{4)  Farid Model:} According to \cite{Farid07}, the Farid pointing errors model is expressed as
\begin{equation}\label{AeqAdd41}
  h_{{p}}(r ; z) = A_0 \exp \left(-\frac{2 r^2}{w_{z_{\mathrm{eq}}}^2}\right),
\end{equation}
where $v=(\sqrt{\pi} a) /\left(\sqrt{2} w_z\right)$, and
\begin{equation}\label{AeqAdd42}
  A_0=[\operatorname{erf}(v)]^2, \quad w_{z_{\mathrm{eq}}}^2=w_z^2 \frac{\sqrt{\pi} \operatorname{erf}(v)}{2 v \exp \left(-v^2\right)}
\end{equation}
with $A_0$ and $w_{z_{eq}}$ denoting the  fraction of the collected power at $r = 0$ and the equivalent beam width respectively. \indent According to (\ref{AeqAdd41}), $c_1 = A_0$ and $c_2 = 2/w_{z_{eq}}^2$ for the Farid model. \\
 \\
\indent \emph{5)  First Reduced  Vasylyev Model:} According to \cite{Vasylyev12}, the pointing errors model established by Vasylyev is expressed as
\begin{equation}\label{Aeq5}
  h_p\left(r;z\right) = \eta \exp\left[-\left(\frac{r}{R}\right)^\lambda\right],
\end{equation}
where $\lambda$ and $R$ are respectively given by
\begin{equation}\label{Aeq6}
\begin{split}
\lambda&=  8 \frac{R_a^2}{w_z^2} \frac{\exp \left(-4 \frac{R_a^2}{w_z^2}\right) {I}_1\left(4 \frac{R_a^2}{w_z^2}\right)}{1-\exp \left(-4 \frac{R_a^2}{w_z^2}\right) {I}_0\left(4 \frac{R_a^2}{w_z^2}\right)} \\
& \times\left[\ln \left(\frac{2 \eta}{1-\exp \left[-4 \frac{R_a^2}{w_z^2}\right] {I}_0\left(4 \frac{R_a^2}{w_z^2}\right)}\right)\right]^{-1}, \\
R&=R_a\left[\ln \left(\frac{2 \eta}{1-\exp \left(-4 \frac{R_a^2}{w_z^2}\right) {I}_0\left(4 \frac{R_a^2}{w_z^2}\right)}\right)\right]^{-\frac{1}{\lambda}}.
\end{split}
\end{equation}
By using the Taylor series of $\exp\left(\cdot\right)$, $I_0\left(\cdot\right)$, and $I_1\left(\cdot\right)$ at $R_a / w_z = 0$ in \cite{gradshteyn2007}, coefficients $\lambda$ and $R$ are then simplified into
\begin{equation}\label{Aeq7}
\begin{split}
\lim_{R_a /w_z  \rightarrow 0  }\lambda &= 2 \\
 \lim_{R_a /w_z  \rightarrow 0 }R &= R_a \left(2R_a^2 / w_z^2\right)^{-1/2} = w_z/2 
\end{split}
\end{equation}
after some algebraic manipulations. Substituting (\ref{Aeq7}) into (\ref{Aeq5}), the first reduced Vasylyev model is obtained as
\begin{equation}\label{Aeq8}
  h_p\left(r;z\right) = \eta \exp\left(-\frac{2r^2}{w_z^2}\right).
\end{equation}
In this case, $c_1 = \eta$, $c_2 = 2 / w_z^2$.
\subsection{Models for $R_a \gg w_z$.}
\begin{figure}
  \centering
  \includegraphics[width=2in]{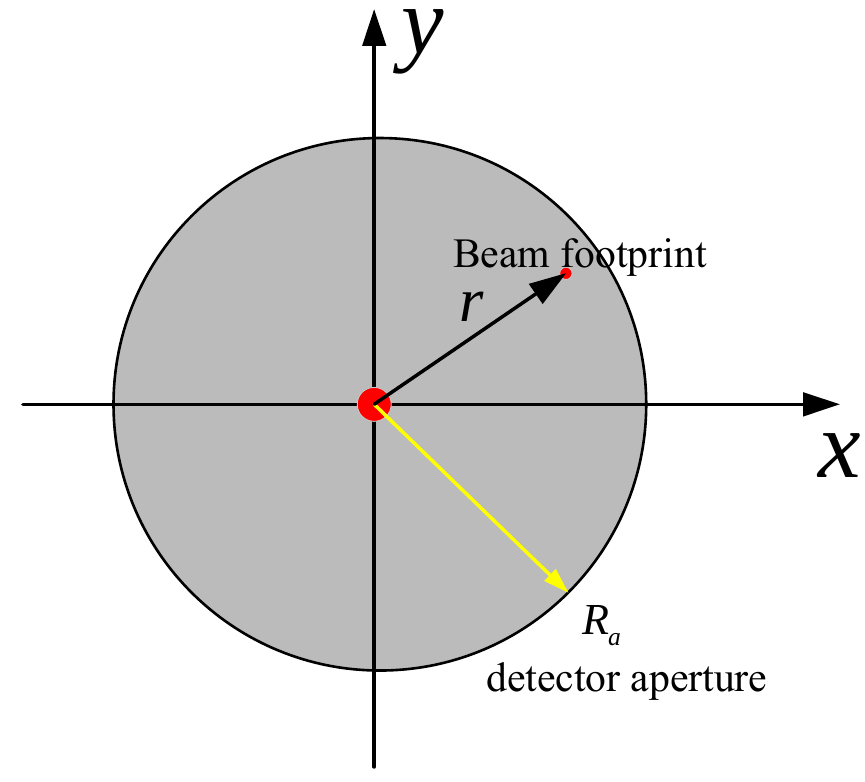}
  \caption{Schematic description of point approximation of beam spot when $R_a \gg w_z$.}\label{Fig4}
\end{figure}
\indent \emph{1)  Point Approximation Model:} In this model, beam spot acts like a \emph{point}, as shown in Fig.~\ref{Fig4}. Note that nearly total energy of laser beam is concentrated around  this point, and this leads to
\begin{equation}\label{Aeq9}
  h_p(r ; z)= \begin{cases}1 & 0 \leq r<R_a \\ 0 & r>R_a\end{cases}.
\end{equation}
\indent Alternatively, the above formula can be rewritten as
\begin{equation}\label{Aeq10}
  h_p(r ; z)= \varepsilon(r)-\varepsilon\left(r-R_a\right)
\end{equation}
where $\varepsilon\left(\cdot\right)$ denotes the Heaviside step function. Specifically, the pointing errors at $r = R_a$ can be obtained as
\begin{equation}\label{Aeq11}
 h_p(r ; z) = \int_0^{w_z} \frac{2}{\pi w_z^2} \exp \left(-\frac{2 r^2}{w_z^2}\right) r \pi d r \approx \frac{1}{2}.
\end{equation}
\indent Combining (\ref{Aeq9}) with (\ref{Aeq11}), the  pointing errors model for the case $R_a \gg w_z$ can then be expressed as
\begin{equation}\label{Aeq11Add1}
    h_p(r ; z)= \begin{cases}1 & 0 \leq r<R_a \\  1/2 &  r = R_a \\ 0 & r>R_a\end{cases}.
\end{equation}
 \indent  The problem  is then transformed to find some smooth function that can approximates (\ref{Aeq11Add1}) efficiently. Inspired by the fact that the logistic function is typically used to approximate the step function, and we develop a pointing errors formula as \footnote{More specifically, (\ref{Aeq12}) can be directly constructed from the Fermi-Dirac distribution, which has these properties of (\ref{Aeq11Add1}). The PDF of Fermi-Dirac distribution is $f=\frac{1}{\exp \left((E-\mu) / k_{\mathrm{B}} T\right)+1}=1-\frac{1}{1+\exp \left(-(E-\mu) / k_{\mathrm{B}} T\right)}$ \cite{Kim21}. Thus, (\ref{Aeq12}) can be obtained by substituting $E, \mu, k_{\mathrm{B}} T$ with $\left({r}/{R_a}\right)^{2 k}, 1,1 / \alpha$ respectively.}
\begin{equation}\label{Aeq12}
  h_p(r ; z)\approx 1-\frac{1}{1+\exp \left(-\alpha\left[\left(\frac{r}{R_a}\right)^{2 k}-1\right]\right)},
\end{equation}
where $k \in \mathbb{Z}^{+}$ with $\mathbb{Z}^{+}$ denoting the set of  positive integers, and $\alpha$ represents the logistic growth rate or steepness of the curve. \\
\indent From (\ref{Aeq12}), the derivative of $h_p(r ; z)$ at $r = R_a$ results to a simple formula, which is obtained as
\begin{equation}\label{Aeq13}
  \left.\frac{d h_p(r ; z)}{d r}\right|_{r=R_a}=-\frac{k}{2} \alpha.
\end{equation}
Hence, combining (\ref{Aeq13}) with \cite[eqn.~(D6)]{Vasylyev12}, the relationship between $k$ and $\alpha$ is given by
\begin{equation}\label{Aeq14}
  \alpha=\frac{8 R_a^2}{k w_z^2} \exp \left(-4 \frac{R_a^2}{w_z^2}\right) I_1\left(4 \frac{R_a^2}{w_z^2}\right).
\end{equation}
\indent  In addition, by using the  asymptotic expansion formula of the modified Bessel function $I_n\left(z\right)$  for large $z$ in \cite{Olver54}, i.e., $I_n(n z) \approx (2 \pi n z)^{-\frac{1}{2}} \exp\left({n z}\right)$, and then (\ref{Aeq14}) reduces into
 \begin{equation}\label{Aeq15}
   \alpha = \frac{2 \sqrt{2} R_a}{\sqrt{\pi} k w_z}.
 \end{equation}
 The above formula indicates that the curve drops faster at the midpoint $r = R_a$ as the ratio between detector aperture and beam width becomes larger, and this is   in line with expectations. \\
 \indent Substituting (\ref{Aeq15}) into (\ref{Aeq12}) gives the result of the {point approximation model}. Note that (\ref{Aeq12}) is reduced to
  \begin{equation}\label{Aeq15Add1}
    h_p(r;z) = \exp\left(-\alpha\left[\left(\frac{r}{R_a}\right)^{2 k}-1\right]\right)
  \end{equation}
  as $r > R_a$. Hence, based on the (\ref{eqp1}) and (\ref{Aeq15Add1}), we demonstrate that the parameter $k$  is assigned  to be 1 intuitively, and this can be verified based on the  numerical  results  in the next section. \\
  \indent Furthermore, by using the (\ref{Aeq1Add1}) and (\ref{Aeq12}), the  PDF of $h_p$ in this case is approximated as
  \begin{equation}\label{Aeq15Add1}
    f_{h_p}\left(h_p\right) =   \frac{R_a^2}{2 \sigma_s^2} \frac{1}{\alpha}\left[\frac{h_p}{c(1-h_p)}\right]^{\frac{R_a^2}{2 \alpha \sigma_s^2}} \frac{1}{h_p(1-h_p)},
  \end{equation}
  where $c = \exp\left(\alpha\right)$. \\
 \indent \emph{2)  Second Reduced Vasylyev Model:} As $R_a \gg w_z$, the coefficients $\lambda, R$ in (\ref{Aeq6}) can be reduced into
 \begin{equation}\label{Aeq16}
 \begin{split}
  \lambda &=  \frac{2 \sqrt{2} R_a}{\sqrt{\pi} w_z} \frac{1}{\ln\left(2\right)}, \\
 R & = R_a {\ln\left(2\right)}^{-1/\lambda},
 \end{split}
 \end{equation}
and a full derivation of (\ref{Aeq16}) is present in Appendix A. \\
 \indent Hence, substituting (\ref{Aeq16}) into (\ref{Aeq5}), the second reduced Vasylyev model is obtained as
 \begin{equation}\label{Aeq18}
   h_p\left(r;z\right) = \exp\left[-\left(\frac{r}{R_a}\right)^\lambda  \ln\left(2\right) \right] = 2^{-\left(\frac{r}{R_a}\right)^\lambda}.
 \end{equation}
 \indent Correspondingly, by using the (\ref{Aeq1Add1}) and (\ref{Aeq18}), the PDF of $h_p$ in this case is obtained approximately  as
 \begin{equation}\label{Aeq19}
 \begin{split}
   f_{h_p}\left(h_p\right) &= \frac{R_a^2}{\ln (2)^{2 / \lambda} \sigma_s^2 \lambda h_p} \exp \left[-\frac{R_a^2}{2 \sigma_s^2}\left(\frac{\ln (1 / h_p)}{\ln (2)}\right)^{2 / \lambda}\right]\\
   & \times  \ln (1 / h_p)^{2 / \lambda-1}.
   \end{split}
 \end{equation}
 \section{NUMERICAL RESULTS}
\begin{figure}[htbp]
  \centering
  \includegraphics[width=2.4in]{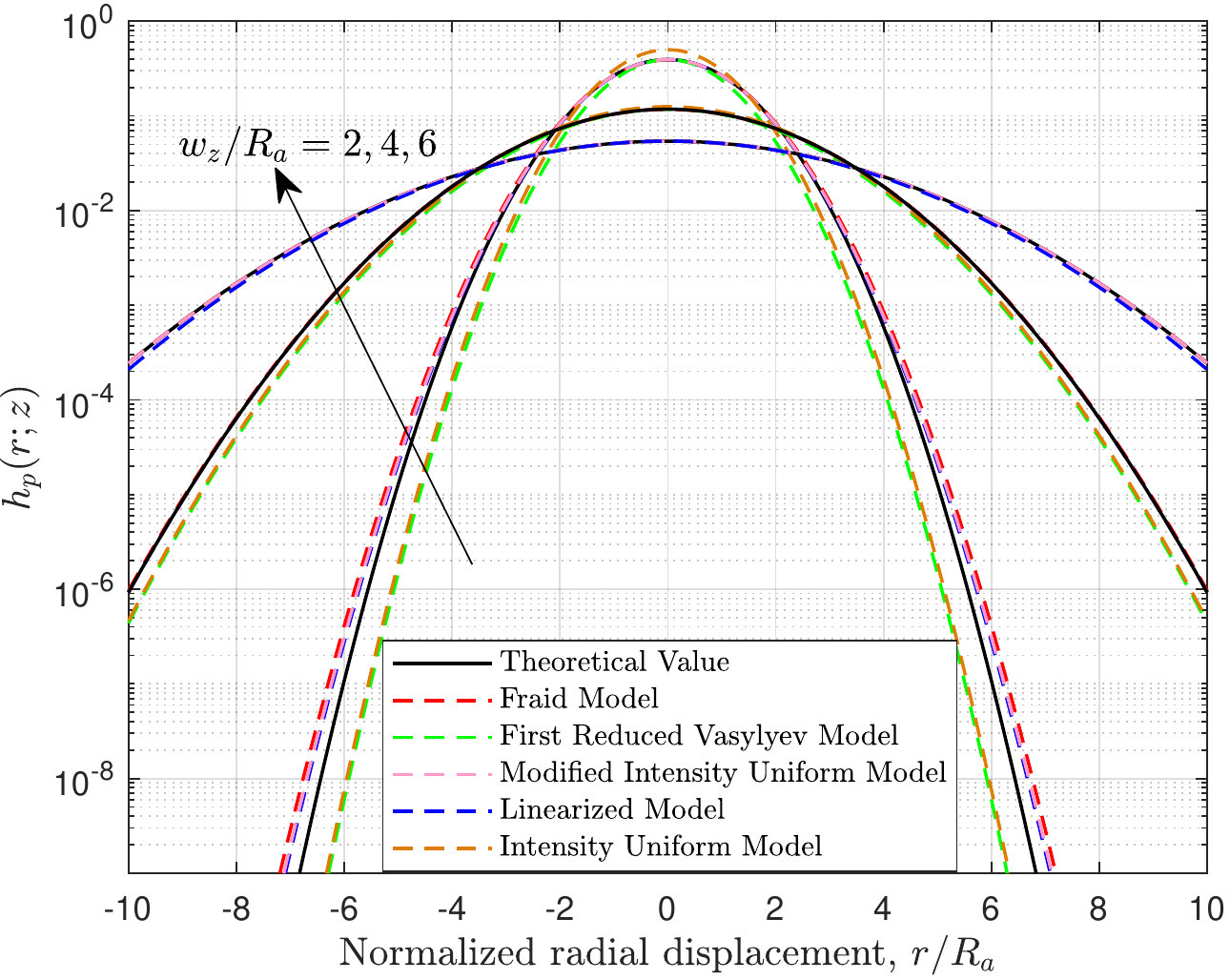}
  \caption{Theoretical and approximate values of $h_p(r;z)$ for different pointing errors models.}\label{Fig5}
\end{figure}

 In this section, we investigate the effectiveness of the pointing errors models that are present in the previous section.  The theoretical results are obtained through MATLAB, and they are also included as a benchmark in all the figures. Moreover, from the perspective of computation efficiency, the number of splits $n$ in the {linearized model} is set to be 4 if not specified yet. It should be emphasized that the radial distance $r_0$ in the {linearized model} is optimized to minimize the NMSE performance, which is defined by $||h-\hat{h}||_2^2 / ||h||_2^2$ with $h$ and $\hat{h}$ representing the theoretical value and approximate value respectively. \\
 \indent In Fig.~\ref{Fig5}, we present the theoretical results and approximate  results for different models and $w_z/R_a$. The corresponding NMSE performance is shown in Table~\ref{tab1}. Note that the normalized optimized radial distance $r_0^*/R_a$ that minimizes the NMSE performance for these three kinds of normalized beam width, i.e., $w_z/R_a = 2, 4, 6$ are $4.05, 12.95, 27.6$ respectively.  From this figure, it can be clearly seen that the accuracy of the {first reduced Vasylyev model} and the {intensity uniform model} is close to each other, and they present  the poorest approximation accuracy among these models.  The accuracy of {modified intensity uniform model} is comparable with that of {linearized model}, and the former is more computation-efficient  than the latter. Moreover,  both of these two models show excellent agreement with the theoretical values even when $w_z/R_a = 2$, where NMSE $\approx 1\times 10^{-5}$, and are more accurate than the traditional Farid model that is widely used in the FSO systems.

\begin{table}[htbp]\small
 \renewcommand\arraystretch{1}
\caption{NMSE PERFORMANCE BETWEEN THEORETICAL VALUES AND APPROXIMATE VALUES FOR DIFFERENT MODELS}
\begin{tabular}{cccc}
  \hline
   Model & $w_z/R_a = 2$ & $w_z/R_a = 4$ & $w_z/R_a = 6$ \\ \hline
  Farid Model & $1.08\times 10^{-4}$ & $5.53\times 10^{-6}$ & $1.13\times 10^{-6}$ \\
  \makecell{  First Reduced \\
  Vasylyev Model} & $1.07\times 10^{-2}$ & $7.17 \times 10^{-4}$ & $1.43\times 10^{-4}$ \\
  \makecell{Modified Intensity \\ Uniform Model} & $1.94\times 10^{-5}$ & $8.95\times 10^{-8}$ & $3.57\times 10^{-9}$ \\
  Linearized Model & $1.14\times 10^{-5}$ & $4.85\times 10^{-8}$ & $1.92\times 10^{-9}$ \\
 \makecell{Intensity Uniform \\ Model} & $4.62\times 10^{-2}$ & $2.74\times 10^{-3}$ & $5.35\times 10^{-4}$ \\
  \hline
\end{tabular}
\label{tab1}
\end{table}
 \begin{figure}[htbp]
  \centering
  \includegraphics[width=2.5in]{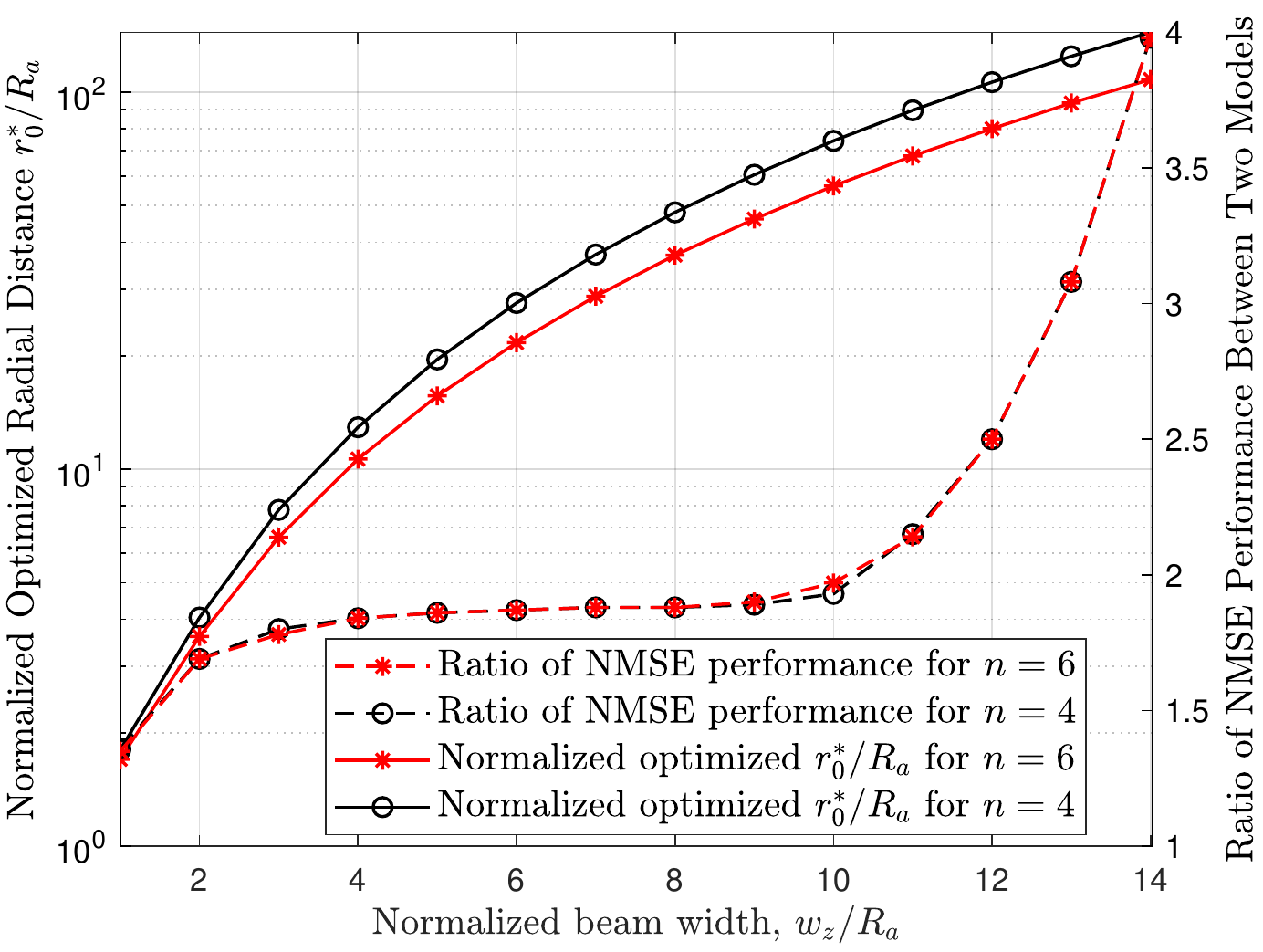}
  \caption{The normalized optimized radial distance $r_0^*/R_a$, and the ratio of NMSE performance between modified intensity
uniform model and linearized model for different $w_z/R_a$.}\label{Fig6}
\end{figure}

 \begin{figure}[t]
  \centering
  \includegraphics[width=2.4in]{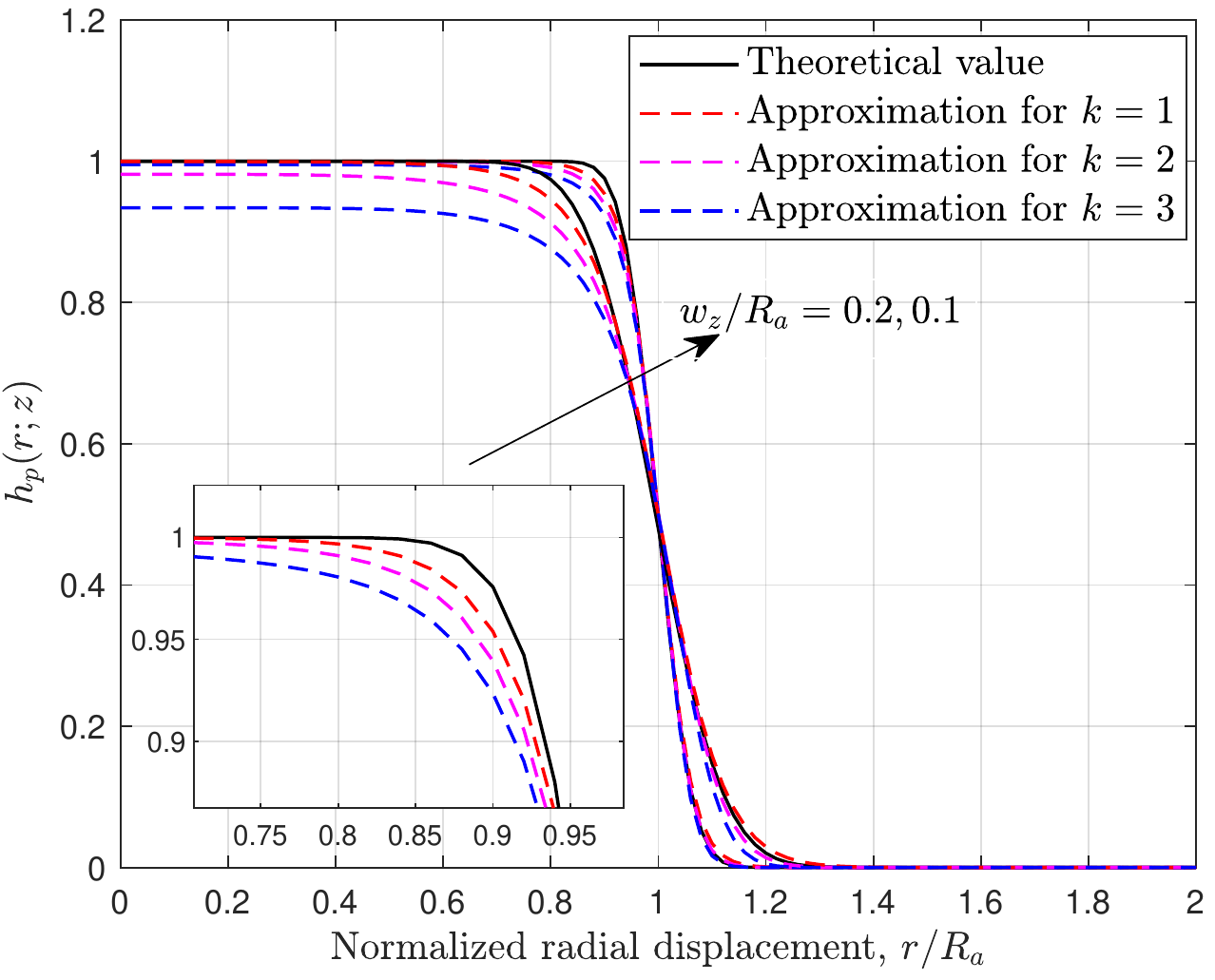}
  \caption{The approximate values of $h_p(r;z)$ for different $k$ in the point approximation model.}\label{Fig7}
\end{figure}
\begin{figure}[t]
  \centering
  \includegraphics[width=2.5in]{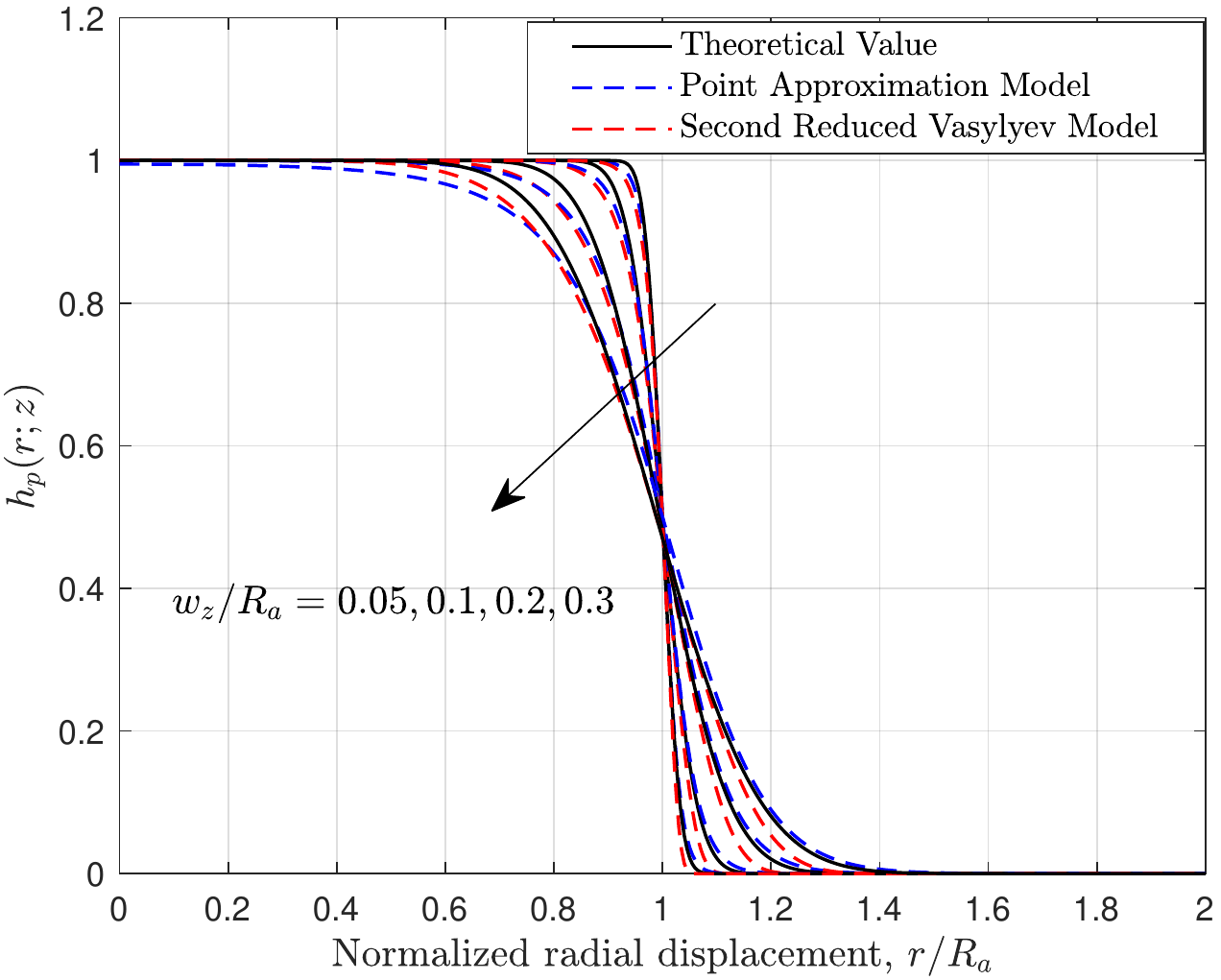}
  \caption{Theoretical and approximate values of $h_p(r;z)$  for different models.}\label{Fig8}
\end{figure}
\begin{table}[b]
\centering
 \renewcommand\arraystretch{1}
\caption{NMSE PERFORMANCE BETWEEN THEORETICAL VALUES AND APPROXIMATE VALUES FOR DIFFERENT MODELS}
\begin{tabular}{ccc}
  \hline

  $w_z/R_a$ & {  Point Approximation
   Model} & \makecell{ Second Reduced \\  Vasylyev Model} \\ \hline
  $0.05$ & $6.05\times 10^{-8}$ & $1.31\times 10^{-5}$ \\
  $0.1$ &  $6.49\times 10^{-7}$ & $3.04\times 10^{-5}$ \\
  $0.2$ & $1.11\times 10^{-5}$ & $4.05\times 10^{-5}$ \\
  $0.3$ & $8.61\times 10^{-5}$ & $1.04\times 10^{-5}$ \\
  \hline
\end{tabular}
\label{tab2}
\end{table}
\indent In Fig.~\ref{Fig6}, we investigate the effects of the normalized beam width $w_z/R_a$, and the number of splits $n$ on the normalized optimized radial distance $r_0^*/R_a$ and the ratio of NMSE performance. It should be noted that the ratio of  NMSE performance is derived between the modified intensity uniform model and the linearized model. From this figure, we find that the relation between $r_0^*/R_a$ and $w_z/R_a$ is  a quadratic function for these two splits, i.e, the expressions  are $r_0^*/R_a = 0.72\left(w_z/R_a\right)^2 + 0.08 \left(w_z/R_a\right)+ 1.01$ and $r_0^*/R_a = 0.52\left(w_z/R_a\right)^2 + 0.30 \left(w_z/R_a\right) + 0.93$ for $n = 4, 6$ respectively. Additionally, the R-square is 1 for both of them. As for the ratio of NMSE performance, it can be observed that  they   are close to each other for two kinds of splits, which  indicates that NMSE performance for $n = 4$ and $n = 6$ is nearly equivalent.  \\
\indent  Fig.~\ref{Fig7} depicts the effects of $k$ in the point approximation model on the approximate accuracy when $w_z/R_a = 0.2$ and $w_z/R_a = 0.1$. Specifically, the corresponding NMSE results for $w_z/R_a = 0.1$ are $5.5\times 10^{-5}, 1.2\times 10^{-4}, 3.1 \times 10^{-4}$ for $k = 1, 2, 3$ respectively. As can be seen, the best approximation can be achieved when $k = 1$. As expected, the curve decreases more dramatically at the midpoint $r/R_a = 1$ when $w_z/R_a$ is smaller.\\
\indent Fig.~\ref{Fig8} shows the approximate results of $h_p(r;z)$ for the point approximation model and the second reduced Vasylyev model. In addition, the corresponding NMSE results of these two models are shown in Table~\ref{tab2}. From this figure, we demonstrate that both of the two models provide an efficient approximation when compared to the theoretical values. However, it can be observed from the NMSE performance in  Table~\ref{tab2} that the proposed point approximation model achieves  a higher accuracy than the second reduced Vasylyev
model when $w_z/R_a \leq0.2$.
\section{CONCLUSION}
\indent In this work, we have present several new results for the pointing errors model, and the accuracy of them is investigated in terms of   NMSE performance. The linearized model was shown to provide the best approximation among these models, and the normalized optimized radial distance $r_0^*/R_a$ in this model has a quadratic relationship with the normalized beam width $w_z/R_a$. Also, we demonstrate that the accuracy of the modified intensity uniform model is not only superior to that of the traditional Farid model from the perspective of the numerical results, but also it is expressed in a simpler form. This indicates that our model is preferable in the  performance analysis of FSO systems  considering the effects of the pointing errors. Furthermore, by  analogizing the beam spot with a point  when $R_a \gg w_z$, the solution of the pointing errors model is transformed to a
smooth function approximation problem, and numerical results show that the proposed pointing approximation model achieves a better approximation than the model developed by Vasylyev  when $w_z/R_a \leq 0.2$.
\section*{APPENDIX A}
\indent  According to \cite[eqn.~(8.451.5)]{gradshteyn2007}, we have
\begin{equation}\label{AppendixA1}
  \lim _{x \rightarrow \infty}\exp\left(-x\right)I_{v}\left(x\right) = \frac{1}{\sqrt{2\pi x}} = 0,
\end{equation}
which leads to 
\begin{equation}\label{AppendixA2}
\begin{split}
1-\exp \left(-4 \frac{R_a^2}{w_z^2}\right) I_0\left(4 \frac{R_a^2}{w_z^2}\right) &= 1, \\
\exp \left(-4 \frac{R_a^2}{w_z^2}\right) I_1\left(4 \frac{R_a^2}{w_z^2}\right) &= \frac{1}{\sqrt{2 \pi 4 \frac{R_a^2}{w_z^2}}}.
\end{split}
\end{equation}
   \indent Therefore, combining with the formula  $\mathop {\lim }\limits_{R_a/w_z \to \infty } \eta = 1-\exp\left(-2R_a^2/w_z^2\right)$, we can obtain
\begin{equation}\label{AppendixA3}
    \begin{split}
\lambda & =\frac{2 \sqrt{2} R_a}{\sqrt{\pi} w_z} \frac{1}{\ln (2)} \\
R & =R_a \ln (2)^{-1 / \lambda}
\end{split}
\end{equation}
after some algebraic manipulations.

\bibliographystyle{IEEEtran}
\bibliography{mybibfile}

\end{document}